\documentclass[aps,prd,reprint]{revtex4-1}
\usepackage{graphicx}% Include figure files
\usepackage{dcolumn}% Align table columns on decimal point
\usepackage{natbib,aas_macros}
\usepackage{color}
\usepackage{amssymb}
\usepackage{amsthm}
\usepackage{mathrsfs}
\usepackage{amsmath}

\newcommand{\msun}{{\rm M_\odot}}

\begin{document}
\title{Trans-Ejecta High-Energy Neutrino Emission from Binary Neutron Star Mergers}

\author{Shigeo S. Kimura$^{1,2,3}$}
\author{Kohta Murase$^{1,2,3,4}$}
\author{Imre Bartos $^{5,6}$}
\author{Kunihito Ioka$^{7}$}
\author{Ik Siong Heng$^{8}$}
\author{Peter M{\'e}sz{\'a}ros$^{1,2,3}$}

\affiliation{$^1$Department of Physics, The Pennsylvania State University, University Park, Pennsylvania 16802, USA}
\affiliation{$^2$Department of Astronomy \& Astrophysics, The Pennsylvania State University, University Park, Pennsylvania 16802, USA}
\affiliation{$^3$Center for Particle and Gravitational Astrophysics, The Pennsylvania State University, University Park, Pennsylvania 16802, USA}
\affiliation{$^4$Yukawa Institute for Theoretical Physics, Kyoto, Kyoto 606-8502 Japan}
\affiliation{$^5$Department of Physics, University of Florida, Gainesville, FL 32611, USA}
\affiliation{$^6$Department of Physics, Columbia University, New York, NY 10027, USA}
\affiliation{$^7$Center for Gravitational Physics, Yukawa Institute for Theoretical Physics, Kyoto, Kyoto 606-8502 Japan}
\affiliation{$^8$SUPA, University of Glasgow, Glasgow G12 8QQ, United Kingdom}

\date{\today}

\begin{abstract}
The observations of a macronova/kilonova accompanied by gravitational waves from a binary neutron star merger (GW170817) confirmed that 
neutron star coalescences produce copious ejecta. The coincident gamma-ray detection implies the existence of a relativistic jet in this system. 
During the jet's propagation within the ejecta, high-energy photons and neutrinos can be produced.
The photons are absorbed by the ejecta, while the neutrinos escape and can be detected. Here, we estimate such trans-ejecta neutrino emission, and discuss how neutrino observations could be used to differentiate between gamma-ray burst emission scenarios. 
We find that neutrinos from the internal shocks inside the ejecta may be detectable by IceCube within a few years of operation, and will likely be detected with IceCube-Gen2. The neutrino signals coincident with gravitational waves would enable us to reveal the physical quantities of the choked jets even without electromagnetic signals. 
\end{abstract}

\maketitle

\section{Introduction}
In 2017, the LIGO and Virgo collaborations reported the detection of gravitational waves (GWs) from a binary neutron star (BNS) merger event, GW170817 \cite{LIGO17c}, which accompanied one of the faintest gamma-ray bursts (GRB) ever observed \cite{LIGO17e,GVB17a}. Electromagnetic signals in other wavelengths were also observed \cite{LIGO17d}: the macronova (also called kilonova) emission in ultraviolet/optical/infrared bands \cite{CBV17a,CFK17a,DPS17Sa,ECK17a,KNS17a,JGEM17a,MASTER17a,ALOP17a,LCO17a,CBV17a,TROS17a,AST17a,MASTER17a,GRAWITA17a,DES17a,DLT4017a} and a slowly brightening afterglow in radio/X-ray bands \cite{HCM17a,HNR17a,MNH18a,RNH18Aa}. 
These electromagnetic counterparts revealed that BNS mergers eject considerable amounts of matter ($\sim 0.01-0.05\msun$; see e.g., Refs. \cite{JGEM17b,WOK17a,MIK18a}) into space, which may choke the relativistic jet launched by the central remnant object \cite{NHS14a,MMR14a,GNP17a,MRM17a}. It is under intense debate whether or not the jet is successful or choked, both of which could explain the prompt gamma rays (see e.g., Refs. \cite{LDM17a,LLC17a,IN17a,VMG18a,LK18a,KBG18a,PBM18a}  for the successful jets and Refs. \cite{KNS17a,2018PhRvD..97h3013S,GNP17a} for the choked jets) and the afterglow (see e.g., Refs. \cite{LPM17a,MAX18a,XZM18a} for the successful jets and Refs. \cite{MNH18a,TPR18a} for the choked jets). Neutrinos and high-energy gamma-rays are not detected in the event \cite{HESS17a,IceCube17c,Fermi17a,SK18a}, although BNS mergers and short GRBs (SGRBs) are considered to be sources of high-energy gamma rays \cite{MTF18a} and neutrinos \cite{KMM17b,BHW18a}. 

If the jet is choked in the macronova/kilonova ejecta, photons from the choked jet are completely absorbed by the ejecta. On the other hand, if neutrinos are produced inside the ejecta (trans-ejecta neutrino), these would be available to look into the inside of the optically thick ejecta. This would enable us to probe the jet physics without electromagnetic signals (e.g., Refs. \cite{MW01a,mi13}). 
Even for the successful jet case, the trans-ejecta neutrinos can be produced when the jet is propagating in the ejecta, which will be observed as a precursor neutrino signals of GRBs.
In this study, we estimate the trans-ejecta neutrino emission of BNS mergers, and discuss the possibility of neutrino detection from the mergers. The neutrino emission at subphotospheres (inside the photosphere) has actively been discussed in the literature of the deaths of massive stars \cite{MW01a,Iocco:2007td,HA08a,Mur08a,Wang:2008zm,MKM13a,mi13,SMM16a,DT18a,HKN18a}. So far this scenario has not been studied in detail in the context of BNS mergers.

This paper is organized as follows. In Section \ref{sec:condition}, we describe the physical situation of the system, and discuss the condition for production of non-thermal protons. Then, we compare important timescales for neutrino production, and estimate the cutoff energies of neutrinos in Section \ref{sec:timescales}. The neutrino signals from internal shocks, detection prospects for future events, and implications for GW170817 are discussed in Section \ref{sec:neutrinos}. We discuss several related issues such as the diffuse neutrino flux in Section \ref{sec:discussion}, and summarize our results in Section \ref{sec:summary}.

\section{Physical Conditions of the System}\label{sec:condition}
\begin{figure}
\includegraphics[width=\linewidth]{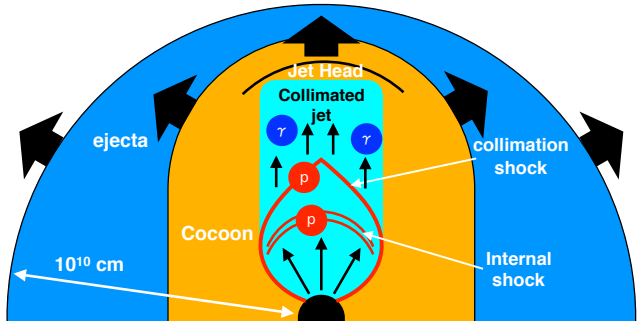}
\caption{Schematic picture of the jet-cocoon system of BNS mergers, where ``p'' and ``$\gamma$'' represent the production site of cosmic-ray protons and target photons.
\label{fig:schematic}}
 \end{figure}

The ejecta of BNS mergers have a few components. One is the dynamical ejecta that consist of the shock-heated and/or tidally stripped material during the merger \cite{HKS13a,BGJ13a}. The remnant object of the merger can be a fast-spinning hyper-massive NS (HMNS) surrounded by a massive accretion torus \cite{KKS14a,STU05a,TRB15a}. Both the HMNS and the accretion torus produce outflows by the viscous and neutrino heating processes \cite{FSK17a,SM17a}. These outflowing material becomes the ejecta of macronova/kilonova of mass 0.01--0.05 $\msun$. The observations of GW170817 suggest two-component ejecta: the fast-blue ($\sim0.3c$) and the slow-red ($\sim0.1-0.2c$) components (see e.g., Refs \cite{KNS17a,JGEM17b,MRK17a}). When the HMNS loses its angular momentum through GW emission and viscosity, it collapses to a black hole, which may lead to the launch of relativistic jets through Blandford-Znajek mechanism \cite{BZ77a,Kom04a,KI15a,TT16a}. The velocity fluctuations of jets make the internal shocks \cite{RM94a}, where the high-energy neutrinos are expected to be produced \cite{WB97a,MN06b}. The jets sweep up the ejecta material during the propagation, forming a cocoon surrounding the jet \cite{RCR02a,ZWM03a,BNP11a,MI13b,NP17a,LDM17a}. If the cocoon pressure is high, it pushes the jet inward, forming a collimation shock. This shock is also likely to produce the high-energy neutrinos \cite{mi13}. In this study, following Ref.~\cite{mi13} for massive stellar collapses, we discuss the neutrino emission from these two sites. Note that we cannot expect particle acceleration at the reverse and forward shocks of the jet head, because the radiation constraint is satisfied there (see Section \ref{sec:mediation}). Figure \ref{fig:schematic} is the schematic picture of this system.

\subsection{Structures of the ejecta and the jet}

We consider a jet propagating in the ejecta of mass $M_{\rm ej}$ and velocity $\beta_{\rm ej}$. We assume a time lag between the ejecta production and the jet launching, $t_{\rm lag}\sim1$ s, and a duration of the jet production similar to that of typical SGRBs, $t_{\rm dur}\sim2$ s. At the time when the jet production stops, the ejecta radius is estimated to be 
\begin{eqnarray}
R_{\rm ej}&=&c \beta_{\rm ej} (t_{\rm dur}+t_{\rm lag})\\
&\simeq& 3.0\times 10^{10} \beta_{\rm ej,-0.48}\chi_{\rm lag,0.18}t_{\rm dur,0.3}\rm~cm ,\nonumber
\end{eqnarray}
where we use $\chi_{\rm lag}= 1 + t_{\rm lag}/t_{\rm dur}$ and notation $Q_x=10^x$ in appropriate unit [$\beta_{\rm ej,-0.48}=\beta_{\rm ej}/(0.33)$, $\chi_{\rm lag,0.18}=\chi_{\rm lag}/1.5$, and $t_{\rm dur,0.3}=t_{\rm dur}/(2\rm~s)$]. Since the fast-blue component is expected to be located in the polar region, we use $\beta_{\rm ej}\simeq 0.33$. This component may originate from the outflow from the HMNS, so we assume the wind-like density profile of the ejecta:
\begin{equation}
\rho_{\rm ej}={M_{\rm ej}\over 4\pi R_{\rm ej}^3} \left({R\over R_{\rm ej}}\right)^{-2}.
\end{equation}
The dynamical ejecta can have a steeper density profile, $\rho_{\rm ej}\propto R^{-3}$, and we do not discuss it for simplicity.
We consider the propagation of the jet whose isotropic equivalent kinetic luminosity $L_{k,\rm iso}$, Lorentz factor $\Gamma_j$, and opening angle $\theta_j$, which leads to the intrinsic jet kinetic luminosity $L_{k,\rm jet}=\theta_j^2L_{k,\rm iso}/2$ (the one-side jet luminosity used in e.g. Refs. \cite{BNP11a,MI13b,HGN17a} is $L_{k,\rm jet}/2$). 
At the downstream of the collimation shock, the jet moves along the jet axis with the Lorentz factor $\Gamma_{\rm cj}\sim\theta_j^{-1}\sim 3.3 \theta_{j,-0.52}^{-1}$ ($\theta_{j,-0.52}=\theta_j/0.3$), which makes the shock Lorentz factor $\Gamma_{\rm rel\mathchar`-cs}\approx \Gamma_j/(2\Gamma_{\rm cj})\simeq 45 \Gamma_{j,2.48}\theta_{j,-0.52}$ ($\Gamma_{j,2.48}=\Gamma_j/300$).  Taking into account the fact that $R_{\rm ej}\propto t$, the jet head position is estimated to be
\begin{eqnarray}
 R_h=2.2\times 10^{10}L_{k,\rm iso,51}^{1/3}\theta_{j,-0.52}^{-2/3}M_{\rm ej,-2}^{-1/3}\\
\times\beta_{\rm ej,-0.48}^{1/3}t_{\rm dur,0.3}^{4/3}\chi_{\rm lag,0.18}^{1/3}\rm~cm, \nonumber
\end{eqnarray}
where $L_{k,\rm iso,51}=L_{k,\rm iso}/(10^{51}\rm~erg~s^{-1})$, $M_{\rm ej,-2}=M_{\rm ej}/(0.01~\msun)$ and we use the fitting formula of Ref. \cite{HGN17a} (see also Ref \cite{MI13b}). This estimate of $R_h$ is at the time of the jet quenching, i.e., $t=t_{\rm dur}$, where $t=0$ is the time when the jet starts being launched. 
The collimation shock forms at 
\begin{equation}
  R_{\rm cs}=9.9\times 10^9L_{k,\rm iso,51}^{1/2}M_{\rm ej,-2}^{-1/2}\beta_{\rm ej,-0.48}^{1/2}t_{\rm dur,0.3}^{3/2}\chi_{\rm lag,0.18}^{1/2}\rm~cm,
\end{equation}
where we use the formula in Ref. \cite{HGN17a} again. 
Note that the pressure gradient that may exist in more realistic situations leads to a collimation shock radius smaller than the estimate above, especially if $R_{\rm cs}\ll R_h$ \cite{MI13b}, although this formula is calibrated to match the results of numerical simulations. 
%The fully analytic formulation given by Ref. \cite{BNP11a} that assumes the constant cocoon pressure. 
In this sense, our setup could be optimistic, since we require that the high-energy neutrino production occurs at radii smaller than $R_{\rm cs}$ as we see later.  
%In this sense, our setup could be optimistic, since we require that the high-energy neutrino production occurs at radii smaller than $R_{\rm cs}$ as we see in the next section. 

For the reference parameter set shown above, $R_h < R_{\rm ej}$ is satisfied at $t=t_{\rm dur}$. This means that the jet is choked before it breaks out from the ejecta, resulting in a dimmer event than the classical SGRBs. The critical luminosity that satisfy $R_h(t_{\rm dur}) = R_{\rm ej}$ is given as 
\begin{eqnarray}
 L_{k,\rm iso,crit}\simeq 2.4\times 10^{51}\theta_{j,-0.52}^{2} M_{\rm ej,-2}\beta_{\rm ej,-0.48}^{2}\label{eq:Lcr}\\
\times t_{\rm dur,0.3}^{-1}\chi_{\rm lag,0.18}^{2} \rm~erg~s^{-1}.\nonumber
\end{eqnarray}
For $L_{k,\rm iso}>L_{\rm iso,crit}$, the jet and the cocoon break out from the ejecta at breakout time $t=t_{\rm bo} < t_{\rm dur}$, resulting in a classical SGRB with a successful jet. For $t<t_{\rm bo}$, the situation is basically the same with the choked jet system, where we can discuss the neutrino emission with the same procedure (see Section \ref{sec:discussion}). For $t>t_{\rm bo}$, our estimate of $R_{\rm cs}$ is no longer valid, so we avoid discussion in detail. 
Note that these estimates assume a wind-like density profile. For the cases with a steeper density profile of the ejecta, $\rho \propto R^{-3}$ as expected for the dynamical ejecta \cite{HKS13a}, the position of the collimation shock and condition for jet breakout are different.
%Note that this situation requires fairly powerful or long-duration jets. 

\begin{figure}
\includegraphics[width=\linewidth]{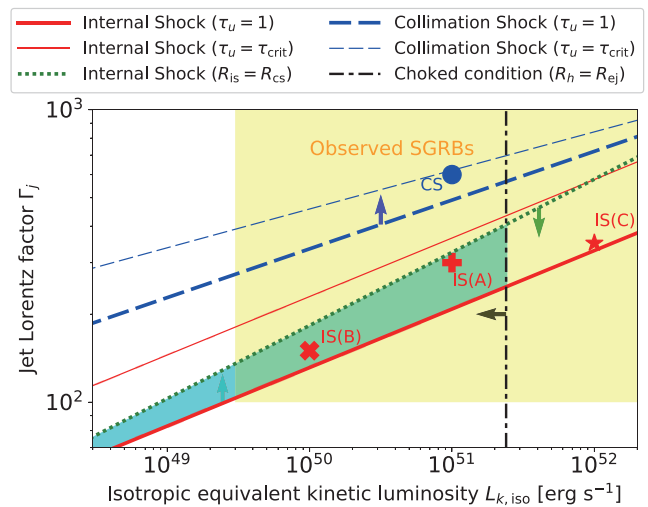}
 \caption{The allowed parameter range on $\Gamma_j$-$L_{k,\rm iso}$ plane for $t_{\rm var}=10^{-4}$ s and $\Gamma_{\rm rel\mathchar`-is}=4$. The radiation constraints in equation (\ref{eq:tau}) are drawn for internal shocks (red-solid lines) and collimation shocks (blue-dashed lines) for $\tau_u < 1$ (thick lines) and $\tau_u < \tau_{\rm crit}$ (thin lines). The dissipation radius condition, $R_{\rm is} < R_{\rm cs}$ (green-dotted line), and the jet breakout condition, $R_h < R_{\rm ej}$ (black-dot-dashed line), are also shown. The allowed parameter region for internal shock models is colored cyan, and the range of observed SGRBs is colored yellow.
 The allowed parameter range for the internal shocks is located on the typical parameter space of SGRBs, while the collimation shock requires higher $\Gamma_j$ to accelerate CRs. 
%Note that the choked jets discussed here can have lower $L_j$ than the observed GRBs who have successful jets.
\label{fig:mediation}}
 \end{figure}

The fluctuations of jet velocity create the internal shocks. The fast shell with the Lorentz factor $\Gamma_r$ catches up the slower one of $\Gamma_s$ at 
\begin{eqnarray}
R_{\rm is}&\approx& 2c t_{\rm var}\Gamma_s^2\approx {c t_{\rm var}\Gamma_j^2\over 2\Gamma_{\rm rel\mathchar`-is}^2}\\
&\simeq& 8.4\times10^9 t_{\rm var,-4}\Gamma_{j,2.48}^2\Gamma_{\rm rel\mathchar`-is,0.6}^{-2}\rm~cm \nonumber,
\end{eqnarray}
where $t_{\rm var}$ is the variability time ($t_{\rm var,-4}=t_{\rm var}/(0.1\rm~ms)$), $\Gamma_j\approx \sqrt{\Gamma_r\Gamma_s}$ is the Lorentz factor of the merged shell, and $\Gamma_{\rm rel\mathchar`-is}\approx \Gamma_r/(2\Gamma_j)$ is the relative Lorentz factor between the merged shell and the fast shell ($\Gamma_{\rm rel\mathchar`-is,0.6}=\Gamma_{\rm rel\mathchar`-is}/4$). Here, we assume that the mass of the fast shell is equal to that of the slow shell, and treat $\Gamma_j$ and $\Gamma_{\rm rel\mathchar`-is}$ as primary parameters. The condition for the internal shock formation in the pre-collimated jet is written as $R_{\rm is}<R_{\rm cs}$, or
\begin{eqnarray}
\Gamma_j < 3.3\times10^2 L_{k,\rm iso,51}^{1/4}M_{\rm ej,-2}^{-1/4}\beta_{\rm ej,-0.48}^{1/4}t_{\rm dur,0.3}^{3/4}\label{eq:breakout_condition}\\
\times \chi_{\rm lag,0.18}^{1/4}t_{\rm var,-4}^{-1/2}\Gamma_{\rm rel\mathchar`-is,0.6},\nonumber 
\end{eqnarray}
The allowed parameter range is shown in Figure \ref{fig:mediation} (green-dotted line).
Note that the internal shocks may be formed in the collimated jet, since the velocity fluctuations exist inside the collimated jet \cite{MI13b}. However, the Lorentz factor in the collimated jet is so low that the internal shocks there cannot avoid being mediated by radiation (see Subsection \ref{sec:mediation}). 
Note also that $t_{\rm var}\sim 0.1$ ms is possible because the dynamical timescale of the central engine is of the order or shorter than it. This short variability timescale can lead to sub-photospheric dissipation, so it would not be observed in canonical GRBs. Also, the GRB analyses with current instruments cannot catch the short variability timescale, since the time bins used in the analyses are longer than a few ms \citep{Swift11a,FermiGBM14a}.    

\begin{table}[tb]%The best place to locate the table environment is directly after its first reference in text
\caption{Model Parameters \label{tab:models}%
}
Shared parameters
\begin{ruledtabular}
\begin{tabular}{ccccc}
$M_{\rm ej} [\msun]$ & $\beta_{\rm ej}$  & $t_{\rm lag}$ [s] & $\theta_j$ & $\xi_B$\\
0.01 & 0.33 & 1 & 0.3 & 0.1 \\
\end{tabular}
\vspace{3pt}
Parameters for the Collimation shock model
\vspace{3pt}
\begin{tabular}{ccccc}
$L_{k,\rm iso} [\rm erg~s^{-1}]$ & $\Gamma_j$& $t_{\rm dur}$ [s] &  $\Gamma_{\rm rel\mathchar`-cs}$ & $\Gamma_{\rm cj}$ \\
10$^{51}$ & 600 & 2 &90  & 3.3\\
\end{tabular}
\vspace{3pt}
Fixed Parameters for the internal shock models
\vspace{3pt}
\begin{tabular}{ccccccc}
$\Gamma_{\rm rel\mathchar`-is}$ & $\epsilon_e$ & $t_{\rm var}$ [s] & $\alpha_1$ & $\alpha_2$  & $\epsilon_p$ & $\xi_{\rm acc}$ \\
4 & 0.1 &  $10^{-4}$ & 0.5 & 2.0 & 0.3 & 1 \\
\end{tabular}
\vspace{3pt}
Parameters for the internal shock models
\vspace{3pt}
\begin{tabular}{ccccc}
model  & $L_{k,\rm iso} [\rm erg~s^{-1}]$ & $\Gamma_j$& $t_{\rm dur}$ or $t_{\rm bo}$ [s] & $\varepsilon_{\gamma,\rm pk}$ [keV] \\
A & 10$^{51}$ & 300 & 2 & 1.7\\
B & 10$^{50}$ & 150 & 2 & 3.3\\
C & 10$^{52}$ & 350 & 0.92 & 1.3\\
\end{tabular}
\end{ruledtabular}
\end{table}

\subsection{Radiation constraints on shock acceleration}\label{sec:mediation}

The non-thermal particle acceleration at the shock requires the sharp velocity change in the gyration scale of the plasma particles, which is achieved if the shock is mediated by the plasma instabilities \cite{Spi08a}. However, when the optical depth of the shock upstream is large, the shock is mediated by radiation, which causes the gradual velocity change in the plasma scale \cite{BKS10a,NS12a}. 
This prevents the particles from being accelerated and gives an important necessary condition for the resulting neutrino emission, as studied in Refs. \cite{mi13,SMM16a}. The condition for the particle acceleration is written using the upstream rest-frame quantities as \cite{mi13,SMM16a}
\begin{equation}
\tau_u = n_u \sigma_T l_u \lesssim 1,~~ {\rm or}~~ \tau_u\lesssim \tau_{\rm crit}\approx {0.1\Gamma_{\rm sh}\over 1+2\ln \Gamma_{\rm sh}^2}\label{eq:tau}
\end{equation}
where $n_u$ is the comoving number density at the shock upstream (hereafter, we use $n$ for the comoving number density and $N$ for the density in the observer frame), $\sigma_T$ is the Thomson cross section, $l_u$ is the length of the upstream fluid, and $\Gamma_{\rm sh}$ is the relative Lorentz factor between the shock upstream and downstream. 
For the non-relativistic flow, the first condition can be used, while the second condition is relevant for the relativistic flow where the electron-positron pairs are produced in the upstream. Since the second condition is derived with assumptions of the high upstream Lorentz factor and the high upstream optical depth, the real acceleration condition for the mildly relativistic flow with a marginal optical depth might lie between the two conditions. Hereafter, we use the former condition for simplicity, although it may be optimistic.

For the collimation shock, the upstream density is written as $n_{\rm cs,u}= L_{k,\rm iso}/(4\pi \Gamma_j^2 R_{\rm cs}^2 m_p c^3)$ and upstream length scale is $l_{\rm cs,u}\approx R_{\rm cs}/\Gamma_j$. Then, the optical depth is estimated to be 
\begin{equation}
 \tau_u\approx 4.4 L_{k,\rm iso,51}^{1/2}M_{\rm ej,-2}^{1/2}\beta_{\rm ej,-0.48}^{-1/2}t_{\rm dur,0.3}^{-3/2}\chi_{\rm lag,0.18}^{-1/2}\Gamma_{j,2.48}^{-3}.
\end{equation}
For the internal shock, the upstream density and length scale are expressed as $n_{\rm is,u}\approx n_{\rm is,d}/\Gamma_{\rm rel\mathchar`-is}$ and $l_{\rm is,u}\approx R_{\rm is}/\Gamma_r$, respectively, where  $n_{\rm is,d}= L_{k,\rm iso}/(4\pi \Gamma_j^2 R_{\rm is}^2 m_p c^3)$ is the density of the merged shell. This leads to 
\begin{eqnarray}
 \tau_u&\approx& {L_{k,\rm iso}\sigma_T\over 8\pi m_p c^3 R_{\rm is} \Gamma_j^3 \Gamma_{\rm rel\mathchar`-is}^2} \nonumber \\ 
&\simeq& 0.16 L_{k,\rm iso,51}R_{\rm is,9.99}^{-1}\Gamma_{j,2.48}^{-3}\Gamma_{\rm rel\mathchar`-is,0.6}^{-2} \\
&\simeq& 0.16 L_{k,\rm iso,51}t_{\rm var,-4}^{-1}\Gamma_{j,2.48}^{-5},\nonumber
\end{eqnarray}
where $R_{\rm is,9.99}=R_{\rm is}/(9.9\times 10^{9} \rm~cm)$. Here, we focus on the reverse shock of the internal shock, since it dissipates more energy. We depict the allowed parameter range for particle acceleration on $\Gamma_j$-$L_{k,\rm iso}$ plane in Figure \ref{fig:mediation}. For our reference parameter set, the collimation shocks are mediated by the radiation unless $\Gamma_j\gtrsim500$, while the internal shocks can be mediated by plasma instabilities for relatively low Lorentz factor $\Gamma_j\gtrsim 200$. Note that the allowed parameter range for the internal shock depends on $\Gamma_{\rm rel\mathchar`-is}$ and $t_{\rm var}$. Higher $\Gamma_{\rm rel\mathchar`-is}$ and/or longer $t_{\rm var}$ make the allowed parameter range (cyan region) larger. 
%We mark the parameter space of typical GRBs by the cyan circle, which have $L_{k,\rm iso}\sim 10^{52}\rm~erg~s^{-1}$ \cite{WP15a} and $\Gamma_j\sim 300$ \cite{GGS13a}. This $\Gamma_j$ is estimated for long GRBs. $\Gamma_j$ for SGRB is not well constrained and usually $\Gamma_j>100$ is assumed. Also, the observed values of $L_j$ for SGRB ranges from $3\times10^{49}\rm~erg~s^{-1}$ to $10^{53}\rm~erg~s^{-1}$. We color this observed parameter range yellow. Note that the choked jets discussed here should have lower $L_j$ than the observed GRBs that are successful jets.

\section{Timescales \&  Critical Energies for Neutrino Production}\label{sec:timescales}

The non-thermal protons produced in the collimation and internal shocks produce neutrinos through interaction with the background photons and protons. Since the physical setup is different, we separately discuss the cases with the collimation shocks and the internal shocks later. Hereafter, we use $\varepsilon_i$ for the particle energy in the comoving frame and $E_i$ in the observer frame.

To calculate the neutrino emission, we need to estimate the cooling and acceleration timescales for protons. 
The acceleration time is estimated to be $t_{\rm acc}\approx \varepsilon_p/(eBc)$, where $B$ is the comoving magnetic field strength.
The cooling timescales are determined by the photomeson production, proton-proton inelastic collision, Bethe-Heitler process, and synchrotron radiation. The photomeson production energy-loss rate is given by
\begin{equation}
 t_{p\gamma}^{-1}=\frac{c}{2\gamma_p^2}\int_{\overline{\varepsilon}_{\rm th}}^\infty{d}\overline{\varepsilon}_\gamma\sigma_{p\gamma}\kappa_{p\gamma}\overline{\varepsilon}_\gamma\int_{\overline\varepsilon_\gamma/(2\gamma_p)}^\infty{d}\varepsilon_\gamma \varepsilon_\gamma^{-2}\frac{dn}{d\varepsilon_\gamma}, \label{eq:pgamma}
\end{equation}
where $\gamma_p=\varepsilon_p/(m_pc^2)$, $\overline\varepsilon_{\rm{th}}\simeq145$~MeV is the threshold energy for the photomeson production, $\overline{\varepsilon}_\gamma$ is the photon energy in the proton rest frame, and $\sigma_{p\gamma}$ and $\kappa_{p\gamma}$ are the cross section and inelasticity for photomeson production, respectively.
We use the fitting formula obtained by Ref. \cite{MN06b} to take the energy dependent $\sigma_{p\gamma}$ and $\kappa_{p\gamma}$ into account.
The timescale of proton-proton interactions is estimated by $t_{pp}^{-1}\approx n_p\sigma_{pp}\kappa_{pp}c$, where $\kappa_{pp}\sim 0.5$ and $\sigma_{pp}\approx (34.3+1.88L+0.25L^2)\times(1-(\varepsilon_{pp,\rm th}/\varepsilon_p)^4)^2$ mb are the inelasticity and the cross section for inelastic proton-proton interaction \cite{kab06}.
We estimate the Bethe-Heitler process cooling rate, $t_{\rm BH}^{-1}$, by Eq. (\ref{eq:pgamma}) using the $\sigma_{\rm BH}$ and $\kappa_{\rm BH}$ instead of $\sigma_{p\gamma}$ and $\kappa_{p\gamma}$. We use the fitting formula in Refs. \cite{SG83a} and \cite{CZS92a} for $\sigma_{\rm BH}$ and $\kappa_{\rm BH}$, respectively. 
The synchrotron timescale for the particle species $i$ is estimated to be  $t_{i,\rm{syn}}=6\pi{m_i^4}c^3/(m_e^2\sigma_TB^2\varepsilon_i)$.

The neutrinos are decay products of pions. The pions also suffer from coolings by the synchrotron and hadronic processes. The hadronic cooling time for pions is approximated to be $t_{\pi p}^{-1}\approx n_p \sigma_{\pi p} \kappa_{\pi p}c$, where $\sigma_{\pi p}\sim 5\times 10^{-26}\rm~cm^2$ and $\kappa_{\pi p}\sim 0.8$ are used for the range of our interest. For the internal shock, adiabatic cooling time $t_{\rm dyn}\approx R_{\rm is}/(c\Gamma_j)$ is also relevant, while it is not applicable for the collimation shock. If the cooling timescale is shorter than the decay time $t_{\pi\rm,dec}=\varepsilon_\pi t_{\pi0}/(m_\pi c^2)$, where $m_\pi$ and $t_{\pi0}$ is the mass and decay time for charged pions at rest, respectively, the neutrino spectrum are suppressed by a factor $f_{\pi,\rm sup}\approx 1-\exp(-t_{\pi\rm,dec}^{-1}/t_{\pi\rm,cl}^{-1})$, where $t_{\pi\rm,cl}^{-1}=t_{\pi p}^{-1}+t_{\pi\rm,syn}^{-1}~(+t_{\rm dyn}^{-1}$ for the internal shock) .  The critical energies at which synchrotron, hadronic, and adiabatic coolings become important are written as $\varepsilon_{\pi,\rm syn}\approx\sqrt{6\pi{m_\pi^5}c^5/(m_e^2\sigma_TB^2t_{\pi0})}$, $\varepsilon_{\pi p}\approx m_\pi c^2/(n_p\sigma_{\pi p}\kappa_{\pi p}c t_{\pi0})$, and $\varepsilon_{\pi,\rm dyn}\approx  m_\pi c^2R_{\rm is}/(2ct_{\pi0}\Gamma_j)$, respectively.
Here, we ignore the inverse Compton cooling of pions for simplicity, which can affect the neutrino spectrum if the target photon energy density is much higher than the magnetic field energy density as included in previous numerical works (see e.g., \cite{Mur08a,mi13}).

\subsection{Collimation shocks}

\begin{figure}
\includegraphics[width=\linewidth]{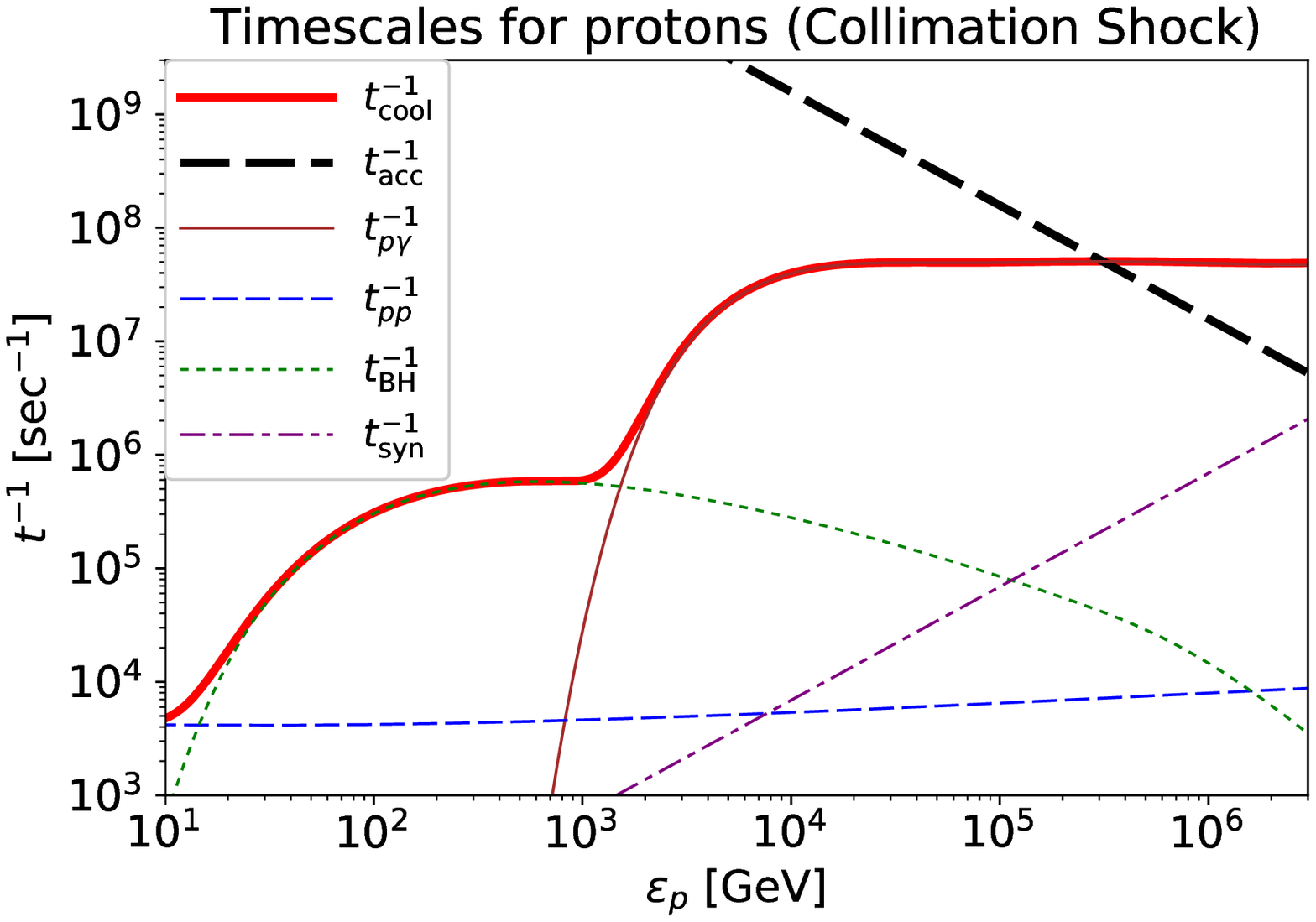}
\includegraphics[width=\linewidth]{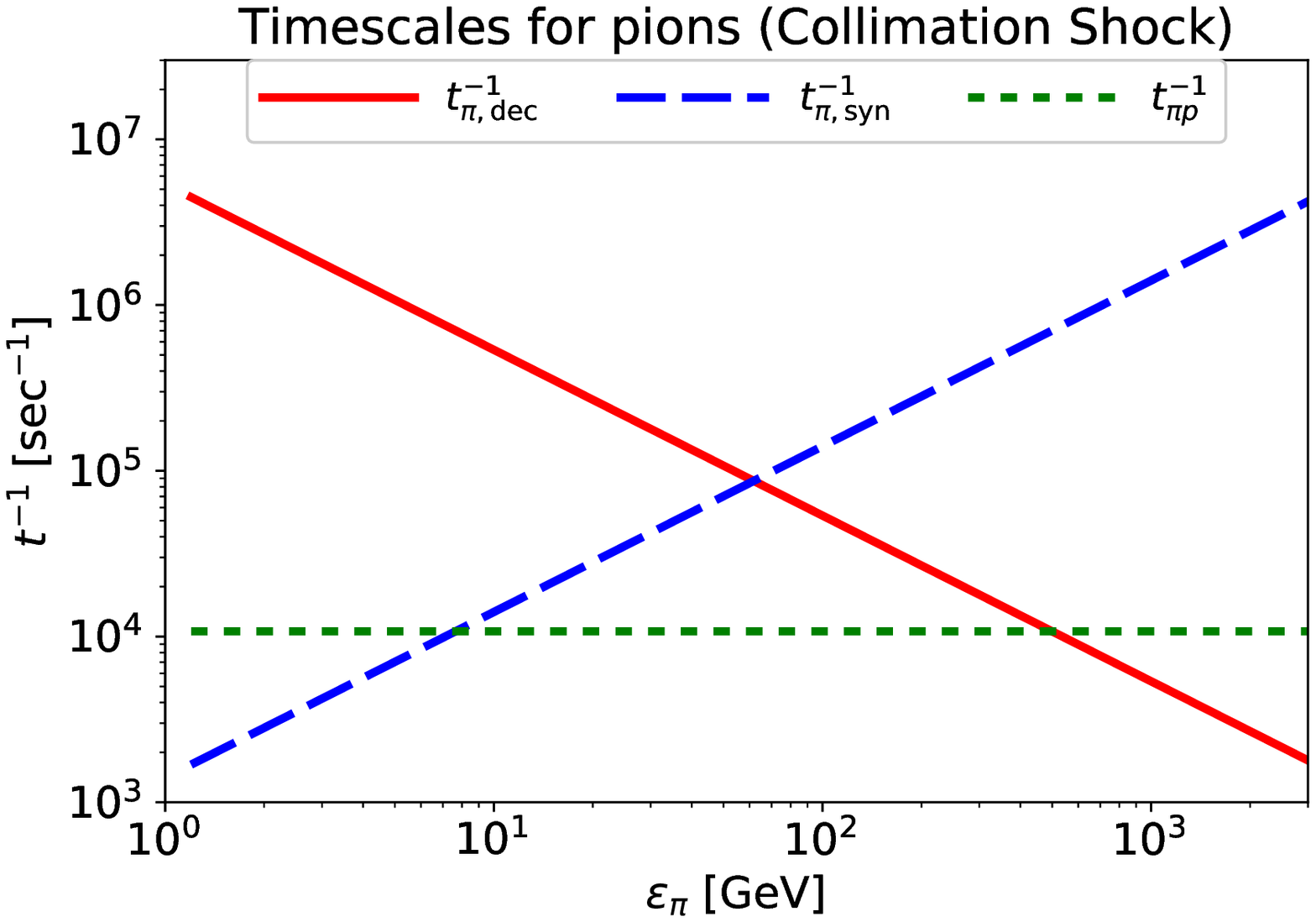}
\caption{The inverse of the timescales at the collimation shock for protons (upper panel) and pions (lower panel) as functions of the comoving proton and pion energy, respectively. 
We can see that in the comoving frame, the maximum energy of protons is $3\times10^2$ TeV, while the pion cooling is effective for $\gtrsim 0.06$ TeV. Since the collimation jets have low Lorentz factor, $\Gamma_{\rm cj}\sim 3$, they produce neutrinos of $E_\nu < $ 0.1 TeV.  
\label{fig:times_CS}}
 \end{figure}

For the collimation shock case, the accelerated protons are advected to the collimated jet, whose density is $n_{\rm cs,d}\approx \Gamma_{\rm rel\mathchar`-cs}n_{\rm cs,u}$. Since the collimated jet is optically thick, we consider the thermal target photons there. Note that the down stream of the collimation shock is always optically thick even if the radiation constraint is avoided at the collimation shock. The radiation constraint is relevant for physical quantities in the upstream of the collimation shock.  Supposing that the thermal energy in the shock downstream are deposited as the radiation, i.e., $U_\gamma=aT^4 \approx (\Gamma_{\rm rel\mathchar`-cs}-1)n_{\rm cs,d}m_pc^2$, the temperature in the collimated jet are estimated to be 
\begin{equation}
T\simeq 9.7 \theta_{j,-0.52}^{1/2} M_{\rm ej,-2}^{1/4}\beta_{\rm ej,-0.48}^{-1/4}t_{\rm dur,0.3}^{-3/4}\chi_{\rm lag,0.18}^{-1/4}\rm~keV. 
\end{equation}
Note that the temperature and radiation energy density in the collimated jet is independent of both $L_{k,\rm iso}$ and $\Gamma_j$. 
%Higher $L_{k,\rm iso}$ makes $R_{\rm cs}$ larger, and higher $\Gamma_j$ makes $n_{\rm is,d}$ lower but $\Gamma_{\rm cs}$ higher.
In the collimated jet, $n_p\approx n_{\rm cs,d}$ and $B\approx \sqrt{8\pi \xi_B aT^4}$, where $\xi_B$ is the ratio of the magnetic field energy density to the radiation energy density. 
%Using this target photons, we estimate $t_{p\gamma}^{-1}$ and $t_{\rm BH}^{-1}$.

We plot the timescales at the collimation shock for protons in the upper panel of Figure \ref{fig:times_CS}, and tabulate the parameters in Table \ref{tab:models}. We do not show another relevant timescale, the advection time $t_{\rm adv}=R_h/(c\Gamma_{\rm cj})$, because it is much longer.
We can see that the Bethe-Heigler process suppresses the pion production for 0.01 TeV $\lesssim \varepsilon_p \lesssim$ 1 TeV, while the pion production efficiency is almost unity above $\varepsilon_p \gtrsim 1$ TeV. The maximum energy of the protons is $\varepsilon_p\simeq 3.1\times10^2$ TeV for our reference parameters. 

The lower panel of Figure \ref{fig:times_CS} show the cooling times and decay time of the pions. We can see that the pion synchrotron is effective for $\varepsilon_\pi \gtrsim$ 0.06 TeV due to the high density and the strong magnetic field in the collimated jet. The critical energies at which synchrotron and hadronic processes become important are estimated to be  $\varepsilon_{\pi,\rm syn}\simeq 0.062  \theta_{j,-0.52}^{-1} M_{\rm ej,-2}^{-1/2}\beta_{\rm ej,-0.48}^{1/2}t_{\rm dur,0.3}^{3/2}\chi_{\rm lag,0.18}^{1/2}\xi_{B,-1}^{-1/2}$ TeV ($\xi_{B,-1}=\xi_B/0.1$) and $\varepsilon_{\pi p}\simeq 0.50 \theta_{j,-0.52}^{-1} \Gamma_{j,2.78} M_{\rm ej,-2}^{-1}\beta_{ej,-0.48} t_{\rm dur,0.3}^3 \chi_{\rm lag,0.18}$ TeV, respectively ($\Gamma_{j,2.78}=\Gamma_j/600$). 
Since the Lorentz factor of the emission region is small, $\Gamma_{\rm cj}\sim 3.3$, the neutrino emission from the collimation shocks will occur at too low energies for detection with IceCube.
%we cannot expect high-energy neutrinos of $E_\nu > 10$ TeV. 
%This makes it difficult to detect the high-energy neutrinos from the collimation shocks near future.

\subsection{Internal shocks}

\begin{figure}
\includegraphics[width=\linewidth]{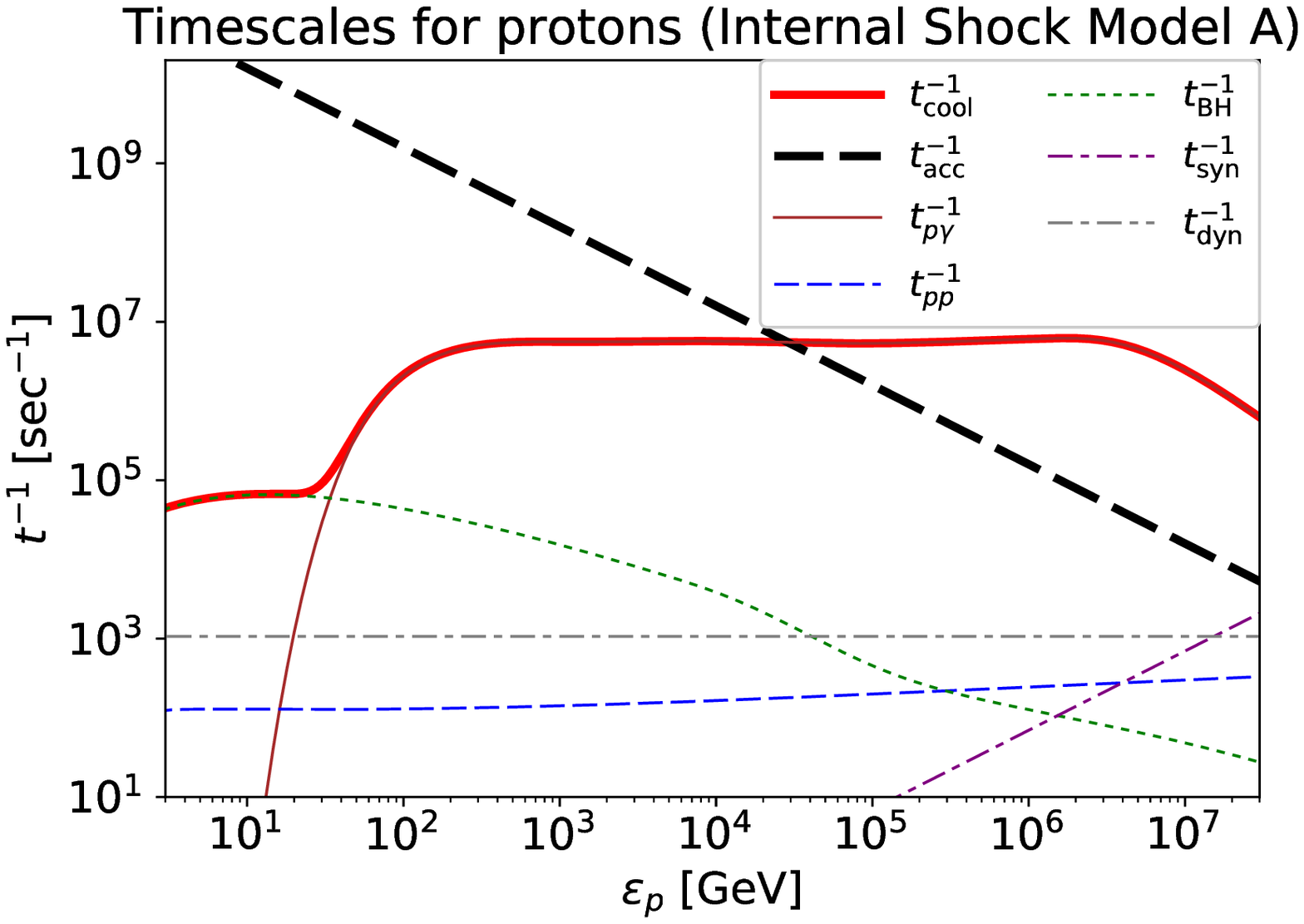}
\includegraphics[width=\linewidth]{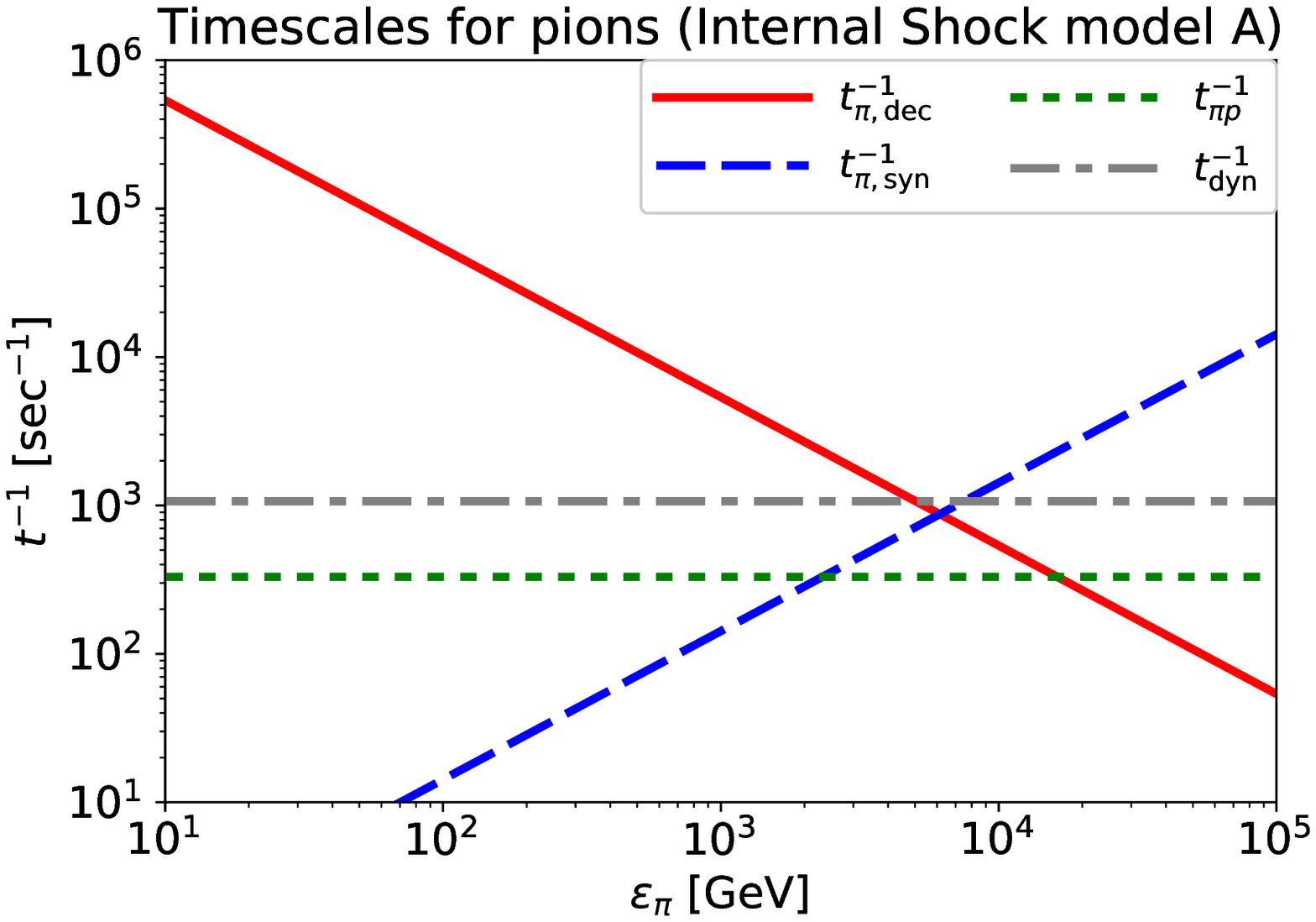}
\caption{Same as Fig. \ref{fig:times_CS}, but for the internal shock with model A. 
The protons are accelerated to a few tens of TeV, and pion cooling is effective around several TeV. The internal shock model has a higher Lorentz factor, $\Gamma_j\sim 300$, so they can emit neutrinos of $E_\nu > 100$ TeV. 
\label{fig:times_IS}}
 \end{figure}

In the internal shocks, we expect two types of the target photons. One is the leakage photons from the collimated jet, and the other is the prompt photons from the non-thermal electrons produced at the internal shock. For the leakage photons, we assume that the escape fraction is $\tau_{\rm cj}^{-1}\sim \Gamma_{\rm cj}/(n_{\rm cs,d}\sigma_TR_{\rm cs})$ \cite{mi13}. Then, the leakage photon density is $\Gamma_j/(2\Gamma_{\rm cj}\tau_{\rm cj})$ times the photon density in the collimated jet, where the factor $\Gamma_j/(2\Gamma_{\rm cj})$ represents the Lorentz boost. The energy of the individual leakage photons is also boosted by $\Gamma_j/(2\Gamma_{\rm cj})$. For the prompt photons, we assume that a fraction $\epsilon_e$ of the thermal energy in the downstream is converted to the non-thermal photon energy, $U_{\gamma,\rm NT}\approx \epsilon_e(\Gamma_{\rm rel\mathchar`-is}-1)n_{\rm is,d} m_p c^2$, and use the broken power-law spectrum, $dn_\gamma/d\varepsilon_\gamma\propto \varepsilon_\gamma^{-\alpha_1}~(\varepsilon_\gamma^{-\alpha_2})$ for $\varepsilon_\gamma < \varepsilon_{\gamma,\rm pk}~(\varepsilon_\gamma > \varepsilon_{\gamma,\rm pk})$. 
The magnetic field at the internal shock is estimated to be $B=\sqrt{8\pi\xi_B U_{\gamma,\rm NT}}$.

In the upper panel of Figure \ref{fig:times_IS}, we plot the inverse of timescales for the internal shock with model A whose parameters are tabulated in Table \ref{tab:models}. The photomeson production is the dominant cooling process in the energy range of our interest, where the contribution from the leakage photons is more important than the prompt photons. Note that these leakage photons have typically higher photon energy, $\varepsilon_\gamma \sim 1-10$ MeV, than the prompt photons, resulting in the high neutrino flux around 1--100 TeV range. The maximum comoving proton energy is estimated to be 30 TeV for model A. 

The pion cooling timescales are shown in the lower panel of the figure. The adiabatic cooling is the most efficient for pions, and the critical energy is 
\begin{eqnarray}
\varepsilon_{\pi,\rm dyn} &\simeq&  5.0 R_{\rm is,9.99}\Gamma_{j,2.48}^{-1} \rm~TeV\nonumber\\
&\simeq& 5.0 t_{\rm var,-4}\Gamma_{j,2.48}\Gamma_{\rm rel\mathchar`-is,0.6}^{-2} \rm~TeV.
\end{eqnarray}
For low $\Gamma_j$ case with fixed $t_{\rm var}$, the hadronic and synchrotron coolings can be important due to their strong $\Gamma_j$ dependence: 
\begin{eqnarray}
&\varepsilon_{\pi,\rm syn}&\simeq 6.1 L_{k,\rm iso,51}^{-1/2}R_{\rm is,9.99}\Gamma_{j,2.48}\Gamma_{\rm rel\mathchar`-is,0.6}^{-1/2}\epsilon_{e,-1}^{-1/2}\xi_{B,-1}^{-1/2} {\rm~TeV}\nonumber \\
&\simeq& 6.1 L_{k,\rm iso,51}^{-1/2}t_{\rm var,-4}\Gamma_{j,2.48}^3\Gamma_{\rm rel\mathchar`-is,0.6}^{-5/2}\epsilon_{e,-1}^{-1/2}\xi_{B,-1}^{-1/2} \rm~TeV,\\
&\varepsilon_{\pi p} &\simeq  16 L_{k,\rm iso,51}^{-1}R_{\rm is,9.99}^2\Gamma_{j,2.48}^2 {\rm~TeV} \nonumber\\
&\simeq& 16 L_{k,\rm iso,51}^{-1}t_{\rm var,-4}^2\Gamma_{j,2.48}^6\Gamma_{\rm rel\mathchar`-is,0.6}^{-4}\rm~TeV.
\end{eqnarray}
Since the Lorentz factor at the emission region for the internal shock case is high, $\Gamma_j\sim 300$, we can expect a high neutrino fluence at $E_\nu > 10$ TeV. 
%Since the target photon energy is high, the Kleing-Nishina effect weakens the inverse Compton (IC) cooling of pions for the energy range of our interest. Hence, we ignore the IC coolings.

\section{Trans-ejecta Neutrinos from the Internal Shocks}\label{sec:neutrinos}

\subsection{Neutrino fluences}

Since the collimation shocks produce lower energy neutrinos that are not suitable for detection by IceCube, we focus on the neutrino emissions from the internal shocks.
For cosmic rays at the internal shock, we use the approximation that a fraction $\epsilon_p$ of the thermal energy at the downstream is deposited on the non-thermal protons. Assuming the canonical shock acceleration spectrum with an exponential cutoff, $dN_p^{\rm iso}/dE_p\propto E_p^{-2}\exp(-E_p/E_{p,\rm max})$, the non-thermal proton spectrum is approximated as
\begin{eqnarray}
E_p^2\frac{dN_p^{\rm iso}}{dE_p}&\approx& \frac{\epsilon_p(\Gamma_{\rm rel\mathchar`-is}-1)\mathscr{E}^{\rm iso}_k}{\ln(E_{p,\rm max}/E_{p,\rm min})} \exp\left(-{E_p\over E_{p,\rm max}}\right)\nonumber \\
&\approx& \frac{\xi_{\rm acc}\mathscr{E}^{\rm iso}_{\rm rad}}{\ln(E_{p,\rm max}/E_{p,\rm min})} \exp\left(-{E_p\over E_{p,\rm max}}\right),\label{eq:Ep2dNdEp}
%E_p^2\frac{dN}{dE_p}
\end{eqnarray}
where $\mathscr{E}^{\rm iso}_k\approx L_{k,\rm iso}t_{\rm dur}$ is the isotropic equivalent kinetic energy, $\xi_{\rm acc}$ is the barion loading factor, $\mathscr{E}^{\rm iso}_{\rm rad}$ is the isotropic equivalent radiation energy, $E_{p,\rm max}$ and $E_{p,\rm min}$ are the maximum and minimum energy of the non-thermal protons at the observer frame, respectively. To convert $\epsilon_p$ and $\mathscr E^{\rm iso}_k$ to $\xi_{\rm acc}$ and $\mathscr{E}^{\rm iso}_{\rm rad}$, we use $\xi_{\rm acc}\approx \epsilon_p/\epsilon_{\rm rad}$ and $\mathscr{E}^{\rm iso}_{\rm rad}\approx \epsilon_{\rm rad} (\Gamma_{\rm rel\mathchar`-is}-1)\mathscr E^{\rm iso}_k$. We use $E_{p,\rm min}\approx \Gamma_j\Gamma_{\rm rel\mathchar`-is} m_p c^2 $ and $E_{p,\rm max}=\Gamma_j\varepsilon_{p,\rm max}$ is obtained by the balance between the acceleration and cooling, i.e., $t_{p,\rm acc}\approx t_{p,\rm cl}$. 
In this work, we set $\epsilon_p=0.3$, $\Gamma_{\rm rel\mathchar`-is}=4$, and $\mathscr{E}^{\rm iso}_{\rm rad}\approx \mathscr{E}^{\rm iso}_k$, which results in  $\xi_{\rm acc}\sim 1$. This value of $\epsilon_p$ is consistent with previous particle-in-cell (PIC) simulations (e.g. \cite{SS11a}). To explain ultrahigh-energy cosmic rays (UHECRs) by long GRBs, $\xi_{\rm acc}\gtrsim 10$ is required  (e.g., \cite{min08}). However, this value may be too optimistic for subphotospheric emission, and $\xi_{\rm acc}\sim1-3$ has also been used in the literature (e.g., \cite{Mur08a,mi13,SMM16a}). Note that we cannot constrain $\epsilon_p$ by the observations, since the normalization of the signals also depends on $\Gamma_{\rm rel\mathchar`-is}$ and $\epsilon_{\rm rad}$. 

\begin{figure}
\includegraphics[width=\linewidth]{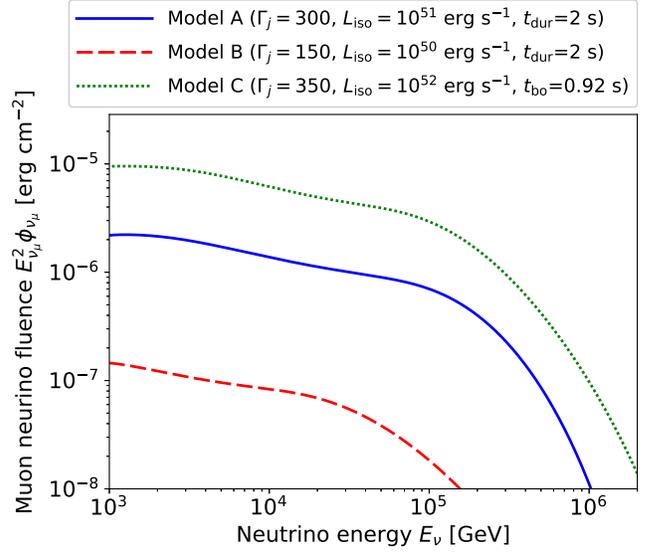}
\caption{The muon neutrino fluences from the internal shock models for optimistic (model A: solid line) and moderate (model B: dashed line) cases for an on-axis observer with $d_L=300$ Mpc. The precursor neutrino fluence from the successful jet (model C: dotted line) is also shown.
Using these fluences, future prospects for coincident detection with gravitational waves are evaluated, which are summarized in Table \ref{tab:prob}. The neutrinos from the collimation shocks are not shown because they are at too low energy.
\label{fig:fluence}}
 \end{figure}

These protons produce pions that decay to muons and muon neutrinos. The muon neutrino spectrum by pion decay is expressed as 
\begin{equation}
E_{\nu_\mu^\pi}^2\frac{dN_{\nu_\mu^\pi}^{\rm iso}}{dE_{\nu_\mu^\pi}}\approx\left(\frac18f_{p\gamma}+\frac16f_{pp}\right)f_{\pi,\rm sup}{E_p^2}\frac{dN_p^{\rm iso}}{dE_p}\label{eq:nu_mu},
\end{equation}
where $f_{p\gamma}=t_{p\gamma}^{-1}/t_{p,\rm cl}^{-1}$ and $f_{pp}=t_{pp}^{-1}/t_{p,\rm cl}^{-1}$ are the neutrino production efficiency through photomeson production and inelastic {\it pp} collision, respectively, and the subscript $\nu_\mu^\pi$ indicates the muon neutrinos produced from pions.
The muons decay to neutrinos and electrons/positrons, whose spectrum is represented as
\begin{equation}
 E_{\nu_e}^2\frac{dN_{\nu_e}^{\rm iso}}{dE_{\nu_e}}\approx E_{\nu_\mu^\mu}^2\frac{dN_{\nu_\mu^\mu}^{\rm iso}}{dE_{\nu_\mu^\mu}}\approx f_{\mu,\rm sup}E_{\nu_\mu^\pi}^2\frac{dN_{\nu_\mu^\pi}^{\rm iso}}{dE_{\nu_\mu^\pi}},%\left(\frac{1}{8}f_{p\gamma}+\frac16f_{pp}\right)f_{\pi,\rm sup}f_{\mu,\rm sup}E_p^2\frac{dN_p}{dE_p},\label{eq:nu_e}
\end{equation}
where $f_{\mu,\rm sup}=1-\exp(-t_{\mu,\rm dec}^{-1}/t_{\mu,\rm cl}^{-1})$ is the suppression factor by the muon cooling, $t_{\mu,\rm cl}^{-1}=t_{\mu\rm,syn}^{-1}+t_{\rm dyn}^{-1}$, and the subscript $\nu_\mu^\mu$ indicates the muon neutrinos produced from muons. 
These muon neutrinos and electron neutrinos change their flavor during the propagation to the Earth. The electron neutrinos and muon neutrino fluences at the Earth are estimated to be~\cite[e.g.,][]{HPS02a}
\begin{equation}
 \phi_{\nu_e+\overline\nu_e}=\frac{10}{18}\phi_{\nu_e+\overline\nu_e}^0+\frac{4}{18}(\phi_{\nu_\mu+\overline\nu_\mu}^0+\phi_{\nu_\tau+\overline\nu_\tau}^0),
\end{equation}
\begin{equation}
 \phi_{\nu_\mu+\overline\nu_\mu}=\frac{4}{18}\phi_{\nu_e+\overline\nu_e}^0+\frac{7}{18}(\phi_{\nu_\mu+\overline\nu_\mu}^0+\phi_{\nu_\tau+\overline\nu_\tau}^0),
\end{equation}
where $\phi_i^0=(dN^{\rm iso}_i/dE_i)/(4\pi{d_L^2})$ is the neutrino fluence without the oscillation and $d_L$ is the luminosity distance. We set $d_L=300$ Mpc as a reference value, which is the declination-averaged horizon distance for face-on NS-NS merger events for the design sensitivity of the second generation detectors \cite{Sch11a}.

The resultant muon neutrino fluences are shown in Figure \ref{fig:fluence} for optimistic (model A) and moderate (model B) sets of parameters tabulated in Table \ref{tab:models}. These models are different in $L_{k,\rm iso}$ and $\Gamma_j$, which mainly affect the normalization of the fluence and the cutoff energy, respectively. For model A, the neutrino spectrum has a cutoff around $E_\nu\sim$ 200 TeV, while for model B, the spectrum break appears at lower energy, $E_\nu\sim$ 50 TeV, due to the lower $\Gamma_j$. The pion cooling causes the cutoff and the spectral break. The combination of the muon cooling and the neutrino oscillation causes a slightly soft spectrum at 3 TeV $\lesssim E_\nu \lesssim 200$ TeV for model A and at 1 TeV $\lesssim E_\nu \lesssim 50$ TeV for model B.

\subsection{Detection rates}\label{sec:detection}
\begin{table}[t]%The best place to locate the table environment is directly after its first reference in text
\caption{Detection probability of neutrinos by IceCube and IceCube-Gen2  \label{tab:prob}%
}
\begin{ruledtabular}
%Detection Probability for a single event \add{at 40\,Mpc}
Number of detected neutrinos from single event at 40\,Mpc
\begin{tabular}{lccc}
model & IceCube (up+hor) & IceCube (down) & Gen2 (up+hor)\\
%A & 0.11 & 0.40 & 0.093\\
A & 2.0  & 0.16 & 8.7  \\
%A & South & 0.55  & --   \\
%B & 6.2$\times10^{-3}$ & 0.026 & 3.5$\times10^{-4}$
B & 0.11 & 7.0$\times10^{-3}$ & 0.46 \\
%B & South & 0.023 & --
\end{tabular}
\vspace{3pt}
Number of detected neutrinos from single event at 300\,Mpc
\begin{tabular}{lccc}
model & IceCube (up+hor) & IceCube (down) & Gen2 (up+hor)\\
A & 0.035 & 2.9$\times10^{-3}$ & 0.15\\
B & 1.9$\times10^{-3}$ & 1.3$\times10^{-4}$ &8.1$\times10^{-3}$
\end{tabular}
\vspace{3pt}
GW+neutrino detection rate [yr$^{-1}$]
\vspace{3pt}
\begin{tabular}{lcc}
%model & IC  & $P_{3\rm yr}$ (IC)  & $P_{1\rm yr}$ (Gen2) & $P_{3\rm yr}$ (Gen2)  \\
model & IceCube (up+hor+down)  & Gen2 (up+hor) \\
A & 0.38 & 1.2 \\
B & 0.024 & 0.091 \\
%A & 0.38 & 0.76 & 0.88 & 0.998 \\
%B & 0.025 &  0.23 & 0.10 & 0.67 \\
\end{tabular}
\end{ruledtabular}
\end{table}

These neutrinos can be detected by IceCube or IceCube-Gen2 as $\nu_\mu$-induced track events, whose expected event number is estimated to be
\begin{equation}
 \overline{\mathcal{N}_\mu}=\int\phi_\nu{A}_{\rm{eff}}(\delta,~E_\nu)dE_\nu,
\end{equation}
where $A_{\rm{eff}}$ is the effective area. IceCube and IceCube-Gen2 can also detect $\nu_e$s and $\nu_\tau$s as shower events (or cascade events). The angular resolution of shower events is much worse than that of track events. Also, the effective area for the shower events is smaller than the upgoing track events. Thus, we focus on the detectability of $\nu_\mu$-induced track events, although the shower events may be important for the merger events in the southern sky.   %The detection probability of $k$ neutrinos is described by the Poisson distribution, $p_k=\overline{\mathcal N}^k\exp(-\overline{\mathcal N})/k!$, and the probability of more than $k$ neutrino detections is $1-\sum_{i<k}p_k$. 

We use the effective area shown in Ref. \cite{IceCube17b} for IceCube. For IceCube-Gen2, the effective volume can be 10 times larger than that of IceCube \cite{Gen214a}. Hence, we use $10^{2/3}$ times larger $A_{\rm eff}$ than that for IceCube, although it depends on the specific configurations. The threshold energy for the neutrino detection is set to 0.1\,TeV for IceCube and 1 TeV for IceCube-Gen2.  The downgoing events suffer from the atmospheric background.   Although the downgoing events can be used to discuss the detectability with IceCube, $A_{\rm eff}$ for the downgoing events with IceCube-Gen2 is quite uncertain. Thus, we focus on the upgoing+horizontal events that have declination $\delta>-5^\circ$ for IceCube-Gen2. KM3NeT will observe the events in the southern sky \cite{KM3NeT16a}, which will help make coincident detections in the near future. Note that the atmospheric neutrinos are negligible owing to the short duration of $t_{\rm dur}\sim 2$\,s.

We calculate the expected number of detected neutrinos for models A and B for a single event located at 40\,Mpc, which are tabulated in the upper part of Table \ref{tab:prob}. IceCube is likely to detect a coincident neutrino signal for our model A if the source is located on the northern sky ($\delta > -5^\circ$). For our model B, detection for a source in the northern sky is also possible, but not guaranteed. For IceCube-Gen2, detection is probable for the northern sky events. If we put the source at 300 Mpc, neutrino detection from a single event is unlikely with IceCube, while it is possible with IceCube-Gen2 if the optimistic event (model A) occurs at  the northern sky.

We now calculate the joint GW+neutrino detection rate for a population of sources, which we assume to be uniformly distributed in the local universe. Using the neutron star merger rate obtained by LIGO, $\mathcal R\sim 1.5\times10^3\rm~Gpc^{-3}~yr^{-1}$ \cite{LIGO17c}, around 170 merger events happen within 300\,Mpc every year. The fraction of on-axis events is $f_b\sim 0.045\theta_{j,-0.52}^2$, leading to an on-axis merger rate $\mathcal R_0\simeq $4.1 yr$^{-1}$  within the upgoing+horizontal coverage area. 

Supposing that all merger events have the same neutrino luminosity, and assuming that all binary neutron star mergers within 300\,Mpc are detected by GW owing to amplification of GW emission to the face-on direction, we estimate the joint GW+neutrino detection rate for IceCube and IceCube-Gen2. The resultant values are tabulated in the lower part of Table \ref{tab:prob}. For model A, neutrino detection is highly probable already after a few years of operation even with IceCube. For model B, it is not easy to make a coincident detection with IceCube, while the detection is probable with IceCube-Gen2 for several years of operation. Note that we do not consider downgoing events with IceCube-Gen2 to avoid the uncertainty of its effective area.

We note that on the time frame of IceCube-Gen2, LIGO is likely to further improve its sensitivity, potentially significantly increasing the coincident detection rate.

\subsection{Implications for GW170817}
The non-detection of a high-energy neutrino counterpart of GW170817 can put some limits on the parameters of a choked jet. GW170817 was followed by a faint SGRB (GRB170817A), a macronova/kilonova, and a slowly brightening afterglow. The faint gamma rays and the brightening afterglow already ruled out the canonical SGRB viewed from on-axis. Two models have been proposed to explain this peculiar SGRB: a wide-angle cocoon inflated by the choked jet \cite{MNH18a} and a structured jet viewed from off-axis \cite{LPM17a,MAX18a}. 

Estimates on the viewing angle of GW170817 vary between $\theta_v\sim 0^\circ-40^{\circ}$ \cite{LIGO17c,Man18a,2018arXiv180404179F}. Based on these results, we cannot fully rule out that the viewing angle was close to or within the beaming angle of the relativistic outflow. Thus, in the framework of the cocoon model, the electromagnetic signals do not rule out the possibility that we observe a choked jet from on-axis. In this case, we would expect high-energy neutrinos from this system. The detection probability is estimated to be $\overline{\mathcal{N}_\mu}\simeq 0.16$ for the parameter set of model A with $d_L=40$\,Mpc. Unfortunately, GW170817 appeared in the southern sky, where the effective area of IceCube is about 5 times smaller than that in the northern sky at 100\,TeV. Hence, we could not put meaningful limits on the parameters of the choked jet of GW170817. If GW170817 had occurred in the northern sky or KM3NeT had been in operation, the expected number of neutrino events is $\overline{\mathcal{N}_\mu}\sim 0.1-2$ (see Table \ref{tab:prob}). This means that IceCube could have detected a few neutrino signals, or that we could have put some limits on the choked jet parameters, such as $\theta_v > \theta_j$ and/or low $\Gamma_j$ so that the radiation constraint can be satisfied. Thus, using the current and near future high-energy neutrino facilities, it is possible to discuss such astrophysical phenomena without using the electromagnetic signals.

The gamma-ray isotropic equivalent energy of GRB170817A was $\sim 10^{47}$ erg, while the muon neutrino isotropic equivalent energy of model A is $\sim 10^{49}$ erg. Thus, the neutrino fluence can be around two orders of magnitude higher than the gamma-ray fluence.
Considering the fact that GRB170817A was borderline detectable by Fermi-GBM, if we observe choked jet systems at larger distances, we may not be able to observe gamma rays. Even in this case, we may be able to probe the jet physics using the neutrino and GW signals.

\section{Discussion}\label{sec:discussion}

If we observe these events from an off-axis angle, $\theta_v$, the neutrino fluence decreases with $\phi_\nu \propto \tilde{\delta}^2$ for $\theta_j\lesssim \theta_v\lesssim2\theta_j$ and $\phi_\nu \propto \tilde{\delta}^3$ for $\theta_v\gtrsim 2\theta_j$,, where $\tilde \delta \approx \Gamma_j/(1+\Gamma_j^{2}(\theta_v-\theta_j))^{2}$ \cite{IN01a,YIN02a,IN17a}. Since the off-axis fluence is much lower than that for the on-axis event, it is difficult to detect off-axis events even though they are more frequent. The beaming factor is less for slower jets, but this may not help increase the off-axis fluence, because the slower jets cannot produce high-energy neutrinos due to both the radiation constraints (see Section \ref{sec:mediation}) and the pion cooling suppression (see Section \ref{sec:timescales}).

If the jets are powerful enough to satisfy $L_{k,\rm iso} > L_{k,\rm iso,crit}$ at $t=t_{\rm dur}$, the jets are expected to be observed as the classical SGRBs for on-axis observers. 
The prompt neutrinos from successful GRB jets tend to emit higher energy neutrinos, $E_\nu\gtrsim 1-10$ PeV \cite{WB97a,MN06b,KMM17b}, especially for SGRBs that might have the higher jet Lorentz factors \cite{GNG18a}. On the other hand, the trans-ejecta neutrinos have much lower energies, $E_\nu\lesssim 0.3$ PeV as shown in Figure \ref{fig:fluence}. This energy can be lower than those from the choked jets arising from the death of massive stars \cite{mi13,SMM16a} because of the small dissipation radius for BNS mergers.

The successful jet can produce neutrinos when the jet is propagating in the ejecta. These neutrinos can be detected as a precursor signal of canonical SGRBs. For a given $L_{k,\rm iso}$, we can estimate the breakout time $t_{\rm bo}$ using Equation (\ref{eq:Lcr}). Then, the radiation isotropic energy fluence is estimated to be $\mathscr{E}^{\rm iso}_k= L_{k,\rm iso}t_{\rm bo}$. Using the same procedure as in Section \ref{sec:neutrinos}, we estimate the neutrino fluence for the precursor neutrinos with a parameter set of model C tabulated in Table \ref{tab:models}. The resultant spectrum is shown in Figure \ref{fig:fluence}. The spectrum shape is similar to that for model A, and the fluence is higher owing to high $\mathscr{E}^{\rm iso}_k$.  The time lag between the precursor neutrinos and SGRB prompt gamma-rays is estimated to be $t_{\rm pre,\nu}\approx R_{\rm prmpt}/(2\Gamma_j^2 c)\sim 0.01$ s, where we use $R_{\rm prmpt}=10^{15}$ cm and $\Gamma_j=350$ for the estimate. Thus, the precursor neutrinos should be observed almost simultaneously with the prompt gamma-rays. So far, IceCube GRB analyses mainly use the long-GRB data \cite{IceCube12a,IceCube15a,IceCube17b}. A dedicated analysis focusing on the SGRBs will be able to put a limit on the precursor neutrinos from SGRBs. 

The all-flavor diffuse neutrino flux from the trans-ejecta neutrinos of BNS merger events is estimated to be 
\begin{eqnarray}
 &E_\nu^2& \Phi_\nu \sim {c\over 4\pi H_0}\frac38 f_{\rm sup}f_{p\gamma}f_z f_b \mathcal R E_p^2 {dN_p^{\rm iso}\over dE_p}\\
&\sim&3.6\times10^{-9}~{\rm GeV~cm^{-2}~s^{-1}~sr^{-1}} f_{\rm sup}f_{p\gamma}f_{z}\xi_{\rm acc}{\mathscr E}_{\rm rad,51.3}^{\rm iso},\nonumber
\end{eqnarray}
where $f_z$ is the redshift evolution factor, $f_{\rm sup}$ is the suppression factor by the muon and pion coolings, $\mathcal R\simeq1500\rm~Gpc^{-3}~yr^{-1}$ and $f_b\simeq 0.045\theta_{j,-0.52}^2$ are the local merger rate and beaming correction factor defined in Section \ref{sec:detection}, respectively, and $E_p^2dN/dE_p$ is given in eq. (\ref{eq:Ep2dNdEp}). For the parameter set of model A, we obtain $\mathscr{E}^{\rm iso}_k\sim\mathscr{E}^{\rm iso}_{\rm rad}\simeq 2\times10^{51}$ erg and $\ln(E_{p,\rm max}/E_{p,\rm min})\simeq10$. 
Note that $\mathcal R f_b$ is about 10 times larger than the local SGRB rate, $\sim10~{\rm Gpc}^{-3}~{\rm yr}^{-1}$, implying the boost factor due to choked jets, $f_{\rm cho}\sim10$.
According to this crude estimate, the neutrinos from choked jets of BNS mergers can possibly contribute a considerable fraction of the observed diffuse neutrino flux discussed by Refs. \cite{2013PhRvL.111b1103A,2013Sci...342E...1I,ice14,ice15a}. This is different from the conclusion drawn for the contribution from successful jets connected to SGRBs \cite{2015JCAP...09..036T}.
However, we should note that the choked jets from the death of massive stars gives even higher neutrino production fluxes for the same CR loading factor \cite{mi13,SMM16a}, which means that the neutrinos from BNS mergers are likely sub-dominant in the context of the choked jet scenarios. 

We have assumed a proton composition for the neutrino calculations above. Since the neutron star mergers produce the $r$-process elements, the jets can consist of  heavy nuclei. However, in both the collimation and internal shocks, these heavy nuclei in the jet are completely photo-dissociated into protons and neutrons due to the very dense photon field (cf. Refs. \cite{HMI12a,ZMK17a}). As seen in Figures \ref{fig:times_CS} and \ref{fig:times_IS}, all the non-thermal protons lose their energy through the photomeson production, which means that the heavy nuclei cannot survive there because photodisintegration has the larger cross section and the lower threshold energy. Note that the heavy elements are disintegrated only in the shocked jets, and bulk of the ejecta contains a plenty of heavy elements so as to produce a macronova/kilonova.

To avoid the radiation constraint, we have considered high $\Gamma_j$ jets so far. However, since we considered relatively smaller dissipation radii, the jets may not have enough time to accelerate to high $\Gamma_j$, which could lead  both shocks to be in the radiation mediated regime. In this situation, a neutron-proton converter (NPC) acceleration mechanisms may produce the non-thermal particles, owing to the neutron rich environment \cite{KMM13a,MKM13a}. In the collimation shock, the NPC mechanism is not effective, since the protons of $E\gtrsim 10$ GeV lose their energy through the Bethe-Heitler process for $\Gamma_j=10^2$. For $\Gamma_j=50$, the NPC mechanism can produce the non-thermal particles, but the density at the collimated jet is high enough to cool the pions of $\varepsilon_\pi \gtrsim$ 0.083 TeV through hadronic interactions. In the internal shock, the NPC mechanism can work efficiently for $\Gamma_j=10^2$. However, lower $\Gamma_j$ leads to lower $\varepsilon_{\pi p}$. Since lower $\Gamma_j$ allows us to use higher $t_{\rm var}$ without violating the jet breakout condition (eq. [\ref{eq:breakout_condition}]), the critical energy for hadronic interaction can be $\varepsilon_{\pi p} \simeq 2.2 L_{k,\rm iso,51}^{-1}t_{\rm var,-3}^2 \Gamma_{j,100}^6\Gamma_{\rm rel\mathchar`-is,0.6}^{-4}$ TeV, where we use $\Gamma_j=100$ and $t_{\rm var}=1$ ms. Therefore, the NPC mechanism is likely to produce GeV--TeV neutrinos efficiently, as is also the case with the canonical GRBs \cite{KMM13a,MKM13a}. 

\section{Conclusion}\label{sec:summary}
We investigated the detection prospects of high-energy trans-ejecta neutrinos from binary neutron star mergers. We considered the situation in which the jet is choked inside the macronova/kilonova ejecta, where neutrinos are the only available signal to probe the physics of the choked jet. We evaluated the particle acceleration condition for the collimation shocks and the internal shocks inside the ejecta, and found that the non-thermal protons can be accelerated for $\Gamma_j\gtrsim 200$ for the internal shocks and $\Gamma_j\gtrsim 500$ for the collimation shocks (see Figure \ref{fig:mediation} in detail). 

We estimated the time scales and critical energies relevant to neutrino production, and showed that the internal shocks can efficiently produce high-energy neutrinos of $E_\nu>10$\,TeV owing to the high Lorentz factor of the emission region. On the other hand, collimation shocks mainly emit neutrinos of $\lesssim 1$\,TeV due to efficient pion cooling, and are therefore difficult to observe. For a lower $\Gamma_j$ jet, the NPC mechanism could produce GeV--TeV neutrinos.

According to the estimated neutrino fluence, we expect the detection of a few neutrinos with IceCube if a merger event happens at 40 Mpc in the northern sky and the jet is directed toward the earth. Thus, if GW170817 occurred in the northern sky or if we had KM3NeT in operation, neutrino signals could have been detected or limits could have been put on the choked jet parameters, even without electromagnetic signals.
We also estimated the joint detection rate of GWs and neutrinos from a uniformly distributed source population. The joint detection is probable with IceCube within a few years of operation for our optimistic scenario. With IceCube-Gen2, the joint detection is probable even for our moderate case. 
See Table \ref{tab:prob} for the summary of the detection prospects.
Also, we roughly estimated the diffuse neutrino flux from this source population, and find that BNS mergers might represent a significant contribution to the diffuse neutrino flux observed by IceCube. 
These results are obtained with few sets of parameters, although we choose the typical values based on the observations and theoretical expectations. A thorough investigation of the parameter dependence remains as a future work.

%%%%%%%%%%%%%%%%%%%%%%%%%%%%%%%%%%%%%%%%%%%%%%%%%%
%%%%%%%%%%%%%%%%%%%%%%%%%%%%%%%%%%%%%%%%%%%%%%%%%%

\medskip
\begin{acknowledgments}
We acknowledge Francis Halzen and Kazumi Kashiyama for fruitful discussions. 
This work is supported in part by the Alfred P. Sloan Foundation, NSF Grant No. PHY-1620777 (K.M.), JSPS Oversea Research Fellowship, the IGC post-doctoral fellowship program (S.S.K.), KAKENHI No. 18H01215, 17H06357, 17H06362, 17H06131, 26287051, (K.I.) and NASA NNX13AH50G (P.M.).
\end{acknowledgments}

\appendix

%%%%%%%%%%%%%%%%%%%%%%%%%%%%%%%%%%%%%%%%%%%%%%%%%%
%%%%%%%%%%%%%%%%%%%%%%%%%%%%%%%%%%%%%%%%%%%%%%%%%%

\bibliography{ssk}
\bibliographystyle{apsrev4-1}
\end{document}